\documentclass[preprint,3p,12pt]{elsarticle}

\usepackage{amsfonts,amsmath,amssymb,mathrsfs,bbm,bm}

\usepackage{graphicx,subfigure,pst-all,psfrag}

\usepackage{color}

\biboptions{comma,sort&compress}

\journal{Chaos, Solitons \& Fractals}

\begin{document}

\begin{frontmatter}

\title{Discriminating image textures with the multiscale two-dimensional complexity-entropy causality plane}

\author[CIOp,Ingenieria]{Luciano Zunino\corref{mycorrespondingauthor}}
\cortext[mycorrespondingauthor]{Corresponding author}
\ead{lucianoz@ciop.unlp.edu.ar}
\author[Maringa]{Haroldo V. Ribeiro}
\ead{hvr@dfi.uem.br}
\address[CIOp]{Centro de Investigaciones \'Opticas (CONICET La Plata - CIC), C.C. 3, 1897 Gonnet, Argentina}
\address[Ingenieria]{Departamento de Ciencias B\'asicas, Facultad de Ingenier\'ia, Universidad Nacional de La Plata (UNLP), 1900 La Plata, Argentina}
\address[Maringa]{Departamento de F\'isica, Universidade Estadual de Maring\'a, Maring\'a, PR 87020-900, Brazil}

\begin{abstract}
The aim of this paper is to further explore the usefulness of the two-dimensional complexity-entropy causality plane as a texture image descriptor. A multiscale generalization is introduced in order to distinguish between different roughness features of images at small and large spatial scales. Numerically generated two-dimensional structures are initially considered for illustrating basic concepts in a controlled framework. Then, more realistic situations are studied. Obtained results allow us to confirm that intrinsic spatial correlations of images are successfully unveiled by implementing this multiscale symbolic information-theory approach. Consequently, we conclude that the proposed representation space is a versatile and practical tool for identifying, characterizing and discriminating image textures.
\end{abstract}


\end{frontmatter}


\section{Introduction}
\label{Introduction}
The development of complexity measures for two or higher dimensional data has been recognized as a long-standing goal~\cite{feldman2003}. Several approaches were introduced during the last two decades for a quantitative distinction between different types of ordering or pattern in two-dimensional signals, such as images~\cite{andrienko2000,cai2006}. In particular, techniques to detect fractal and multifractal features have been shown to be useful for dealing with the characterization of self-similar and extended self-similar objects~\cite{kaplan1995,kaplan1999,alvarez-ramirez2006,gu2006,carbone2007}. These approaches have their roots in the seminal work of Mandelbrot~\cite{mandelbrot1982}, who just introduced fractal geometry to mimic natural textured patterns. Cloudy textures, such as those associated with mammographic, terrain, fire, dust, cloud, and smoke images can be suitably described by these scaling and multiscaling analysis~\cite{kaplan1999,alvarez-ramirez2006}. Actually, recent effective applications in heterogeneous fields confirm that these fractal techniques are highly valuable tools, \textit{e.g.} identification of lesion regions of crop leaf affected by diseases~\cite{wang2013}, Hurst exponent estimation performed on satellite images to measure changes on the Earth's surface~\cite{valdiviezo2014}, and determination of scaling properties in encrypted images~\cite{vargas2015}. Despite all these significant efforts, the development of a robust methodology to detect and quantify spatial structures in images still represents an open and subtle problem. Along this research direction, we have previously introduced an extension of the complexity-entropy causality plane to more than one dimension~\cite{ribeiro2012}. It has been shown that the two-dimensional version of this information-theory-derived tool is very promising for distinguishing between two-dimensional patterns. Motivated by this fact, in the present paper, we implement the two-dimensional complexity-entropy causality plane in different numerical and experimental contexts with the aim of testing its potentiality as a texture image quantifier. Furthermore, a multiscale generalization of the original recipe is proposed for characterizing the dominant textures at different spatial scales. As it will be shown, this multiscale approach offers a considerable improvement to the original tool introduced in Ref.~\cite{ribeiro2012}. Since any image corresponds to a two-dimensional ordered array, we conjecture that the proposed multiscale ordinal symbolic approach can be a useful alternative for an efficient and robust characterization of its features, offering deeper insights into the understanding of the underlying phenomenon that governs the spatial dynamics of the system at different resolution scales.

This paper is organized as follows. In the next section, a brief review of the two-dimensional complexity-entropy causality plane is given. Besides, its generalization to multiple spatial scales is also described. In Section~\ref{Results}, we have included several numerical and experimental applications. More precisely, in Section~\ref{2D_ornaments}, an initial periodic ornament is carefully analyzed when adding a variable degree of randomness by changing the color of each pixel with a given probability. A numerically controlled example to illustrate the importance of implementing a multiscale analysis is detailed in Section~\ref{Multiscale_analysis}. The normalized Brodatz texture database is studied in Section~\ref{Brodatz} and results obtained from the characterization of some real images of interest are included in Section~\ref{Experimental}. Finally, the main conclusions of this research are summarized in the last Section~\ref{Conclusions}.

\section{Complexity-entropy causality plane for two-dimensional patterns}
\label{CECP-2D}
A two-dimensional symbolization procedure, following the encoding scheme introduced by Bandt and Pompe (BP)~\cite{bandt2002}, is applied to the image under study. Given a $N_x \times N_y$ image (2D array), the symbolic sequences are obtained by considering the spatial ranking information (ordinal or permutation patterns) associated with overlapping subarrays of size $D_x \times D_y$. This procedure can be better introduced with a simple example; let us assume that we start with the $3 \times 3$ array given below
$$
A=
\left(\begin{array}{ccc}
3 & 4 & 8 \\
5 & 6 & 7 \\
2 & 8 & 9 \end{array}\right).
$$
Four parameters, the embedding dimensions $D_x, D_y \ge 2$ ($D_x,D_y \in \mathbb{N}$, the number of symbols that form the ordinal pattern in the two orthogonal directions) and the embedding delays $\tau_x$ and $\tau_y$ ($\tau_x,\tau_y \in \mathbb{N}$, the spatial separation between symbols in the two orthogonal directions) are chosen. The matrix is partitioned into overlapping subarrays of size $D_x \times D_y$ with delays $\tau_x$ and $\tau_x$ in the horizontal and vertical directions, respectively. The elements in each new partition are replaced by their ranks in the subset. For instance, if we set $D_x=D_y=2$ and $\tau_x=\tau_y=1$, there are four different partitions associated with $A$. The first subarray $A_1=\left(\begin{array}{cc} a_0 & a_1 \\ a_2 & a_3 \end{array}\right)=\left(\begin{array}{cc} 3 & 4 \\ 5 & 6 \end{array}\right)$ is mapped to the ordinal pattern $(0123)$ since $a_0 \le a_1 \le a_2 \le a_3$. The second partition is $A_2=\left(\begin{array}{cc} a_0 & a_1 \\ a_2 & a_3 \end{array}\right)=\left(\begin{array}{cc} 4 & 8 \\ 6 & 7 \end{array}\right)$, and $(0231)$ will be its related ordinal motif because $a_0 \le a_2 \le a_3 \le a_1$. The next subarray $A_3=\left(\begin{array}{cc} a_0 & a_1 \\ a_2 & a_3 \end{array}\right)=\left(\begin{array}{cc} 5 & 6 \\ 2 & 8 \end{array}\right)$ is associated with the ordinal pattern $(2013)$, and the last one $A_4=\left(\begin{array}{cc} a_0 & a_1 \\ a_2 & a_3 \end{array}\right)=\left(\begin{array}{cc} 6 & 7 \\ 8 & 9 \end{array}\right)$ is also mapped to the motif $(0123)$. Subarrays with consecutive elements are taken in the above example because the embedding delays $\tau_x$ and $\tau_y$ are fixed equal to one. However, non-consecutive elements of the original array can be considered by changing the embedding delays. For instance, by choosing $\tau_x=2$ and $\tau_y=1$ only two partitions are obtained from array A, namely $A_1=\left(\begin{array}{cc} 3 & 8 \\ 5 & 7 \end{array}\right)$ and $A_2=\left(\begin{array}{cc} 5 & 7 \\ 2 & 9 \end{array}\right)$. Their related ordinal permutations will be $(0231)$ and $(2013)$, respectively. Finally, in the case $\tau_x=\tau_y=2$ only one subarray, $A_1=\left(\begin{array}{cc} 3 & 8 \\ 2 & 9 \end{array}\right)$, with motif $(2013)$ can be obtained. It is worth remarking that different spatial resolution scales are taken into account by changing the embedding delays.

An ordinal pattern probability distribution 
\begin{equation}
P_{BP}=\{p(\pi_i),i=1,\dots,(D_x D_y)!\},
\label{Ord_prob}
\end{equation}
is subsequently obtained by computing the relative frequencies of the $(D_x D_y)!$ possible ordinal patterns $\pi_i$. For a reliable estimation of this distribution, the image size must be at least an order of magnitude larger than the number of possible ordinal states, \textit{i.e.} $N_x N_y \gg (D_x D_y)!$. It is clear that the 2D encoding scheme previously described is not univocally defined. Actually, instead of ordering the elements row-by-row, an alternative column-by-column ordering recipe could be proposed. However, the BP probability distribution (Eq.~(\ref{Ord_prob})) would remain unchanged since only the label given to the accessible states would change by implementing this alternative definition.

Once the BP probability distribution has been obtained, any information-theory-derived quantifier can be estimated. In particular, and in order to introduce the complexity-entropy diagram, the two involved measures---entropy and complexity---need to be defined. Around a decade ago, Rosso~\textit{et al.}~\cite{rosso2007} proposed to use the normalized Shannon entropy and the normalized Jensen-Shannon complexity for such a purpose. It has been shown that chaotic and stochastic time series are located at different regions of this representation space, thus allowing an efficient discrimination between these two kinds of dynamics that are commonly very hard to distinguish. Given any arbitrary discrete probability distribution $P=\{p_i,i=1,\dots, M\}$, the Shannon's logarithmic information measure is given by
\begin{equation}
S[P]=-\sum_{i=1}^{M}~p_i~\ln p_i~.
\label{shannon_entropy}
\end{equation}
The Shannon entropy $S[P]$ is regarded as a measure of the uncertainty associated to the physical processes described by the probability distribution $P$. It is equal to zero when we can predict with full certainty which of the possible outcomes $i$ whose probabilities are given by $p_i$ will actually take place. Our knowledge of the underlying process described by the probability distribution is maximal in this instance. In contrast, this knowledge is minimal and the entropy (ignorance) is maximal ($S_{max}=S[P_e]=\ln M$) for the equiprobable distribution, \textit{i.e.} $P_e=\{p_i=1/M,i=1,\dots, M\}$. The Shannon entropy is a quantifier for randomness. It is well-known, however, that the degree of structure present in a process is not quantified by randomness measures and, consequently, measures of statistical or structural complexity are necessary for a better understanding of complex dynamics~\cite{feldman1998}. As stated by Lange~\textit{et al.}~\cite{lange2013}: \textit{One would like to have some functional C[P] adequately capturing the ``structurednes'' in the same way as Shannon's entropy captures randomness}. There is not a universally accepted definition of \textit{complexity}~\cite{wackerbauer1994}. In this work, we have implemented the effective statistical complexity measure (SCM) introduced by Lamberti~\textit{et al.}~\cite{lamberti2004}, following the seminal notion advanced by L\'opez-Ruiz~\textit{et al.}~\cite{lopezruiz1995}, through the product
\begin{equation}
\mathcal{C}_{JS}[P]=\mathcal{Q}_{J}[P,P_e] \ \mathcal{H}_{S}[P]
\label{C_JS}
\end{equation}
of the normalized Shannon entropy
\begin{equation}
\mathcal{H}_{S}[P]=S[P]/S_{max}
\label{normalized_shannon_entropy}
\end{equation}
and the disequilibrium $\mathcal{Q}_{J}[P,P_e]$ defined in terms of the Jensen-Shannon divergence. That is,
\begin{equation}
\mathcal{Q}_{J}[P,P_e]=\mathcal{J}[P,P_e]/\mathcal{J}_{max}
\label{disequilibrium}
\end{equation}
with
\begin{equation}
\mathcal{J}[P,P_e]=S[(P+P_e)/2]-S[P]/2-S[P_e]/2
\label{JS_divergence}
\end{equation}
the above-mentioned Jensen-Shannon divergence and $\mathcal{J}_{max}$ the maximum possible value of $\mathcal{J}[{P,P_e}]$. Being more precise, $\mathcal{J}_{max}=-\frac{1}{2} \left [\frac{M+1}{M} \ln(M+1)-2\ln(2M)+\ln(M) \right ]$ is obtained when one of the components of $P$, say $p_m$, is equal to one and the remaining $p_i$ are equal to zero. The Jensen-Shannon divergence quantifies the difference between two (or more) probability distributions. For further details about this information-theory divergence measure please see Refs.~\cite{Lin1991,grosse2002}. Note that the above introduced SCM depends on two different probability distributions, the one associated to the system under analysis, $P$, and the uniform distribution, $P_e$. Furthermore, it was shown that for a given value of $\mathcal{H}_{S}$, the range of possible $\mathcal{C}_{JS}$ values varies between a minimum $\mathcal{C}_{JS}^{min}$ and a maximum $\mathcal{C}_{JS}^{max}$, restricting the possible values of the SCM in a given complexity-entropy plane~\cite{martin2006}. Thus, it is clear that important additional information related to the correlational structure between the components of the system is provided by evaluating the statistical complexity measure.

In this work, the normalized Shannon entropy, $\mathcal{H}_S$ (Eq.~(\ref{normalized_shannon_entropy})), and the SCM, $\mathcal{C}_{JS}$ (Eq.~(\ref{C_JS})), are evaluated using the permutation probability distribution (Eq.~(\ref{Ord_prob})). Defined in this way, these quantifiers are usually known as \textit{permutation entropy} and \textit{permutation statistical complexity}~\cite{zunino2012}. They characterize the diversity and correlational structure, respectively, of the spatial orderings present in the image. The \textit{complexity-entropy causality plane} (CECP) is defined as the representation space obtained by plotting permutation statistical complexity (vertical axis) versus permutation entropy (horizontal axis) for a given system~\cite{rosso2007}. The term \textit{causality} remembers the fact that spatial correlations between samples are taken into account through the BP recipe used to estimate both information-theory quantifiers. The implementation of this kind of diagram as a diagnostic tool was originally proposed by L\'opez-Ruiz~\textit{et al.}~\cite{lopezruiz1995} more than two decades ago. The large number of applications spread over multiple lines of research is a solid proof of its success. Without being exhaustive, we can mention the classification of literary texts~\cite{rosso2009}, the characterization of several financial time series (particularly records obtained from stock~\cite{zunino2010}, commodity~\cite{zunino2011}, and bond~\cite{zunino2012b,fernandez_bariviera2013} markets), the discrimination of music genres~\cite{ribeiro2012b}, the identification of a universal behavior in the complex dynamics of x-ray astrophysical sources~\cite{lovallo2011}, the analysis of stream flow time series within hydrological studies~\cite{lange2013,serinaldi2014}, the description of brain development in chickens~\cite{montani2014}, the quantification of nonstationarity effect in boundary-layer vertical velocity time series~\cite{li2014}, and the study of fluctuating time series of different turbulent plasmas~\cite{weck2015}. 

As it has been mentioned before, the embedding delays determine the spatial separation between symbols. That is, they physically correspond to multiples of the spatial resolution scale of the image under analysis. Consequently, different spatial resolution scales can be scanned by changing the embedding delays of the symbolic reconstruction. We propose to generalize the estimation of both symbolic quantifiers (permutation entropy and statistical complexity) to different embedding delays $\tau_x$ and $\tau_y$ for given embedding dimensions $D_x$ and $D_y$. This \textit{multiscale} CECP seems to be particularly suitable for characterizing the spatial correlations of images at different resolution scales.

\section{Numerical and experimental results}
\label{Results}
 
\subsection{Two-dimensional noisy ornaments}
\label{2D_ornaments}
\begin{figure}
\begin{center}
\resizebox{1.00\columnwidth}{!}{\includegraphics{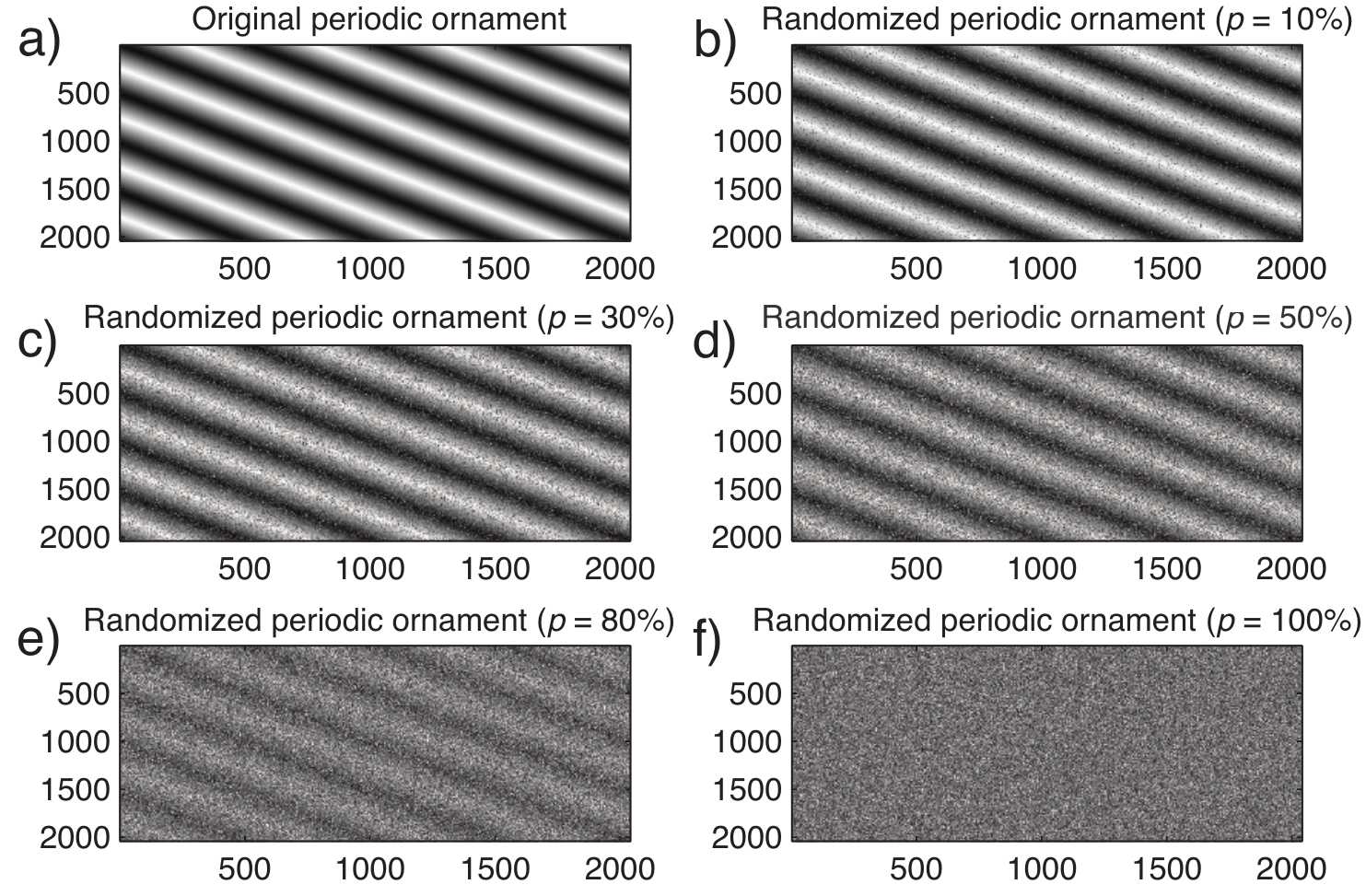}}
\caption{The original periodic image is depicted in a). Randomized periodic ornaments with probabilities $p=10\%$, $p=30\%$, $p=50\%$, $p=80\%$ and $p=100\%$ are illustrated in b), c), d), e) and f), respectively. Images are composed of 256 grayscale levels and have $2,040 \times 2,040$ pixels.}
\label{figure1}
\end{center}
\end{figure}

As a first numerically controlled application and following the geometric ornament model proposed in Refs.~\cite{andrienko2000,cai2006}, we have generated a periodic figure composed of 256 grayscale levels of size $2,040 \times 2,040$. The original periodic ornament is shown in Fig.~\ref{figure1}~a). This very regular 2D array is gradually randomized by changing the color of each pixel with probability $p$. An aleatory grayscale level is assigned to the modified pixels. In Figs.~\ref{figure1}~b)-f) five randomized ornaments are illustrated. Spatial correlations are clearly dominant for lower values of $p$ whereas more random patterns emerge for higher values of this parameter.

\begin{figure}
\begin{center}
\resizebox{1.00\columnwidth}{!}{\includegraphics{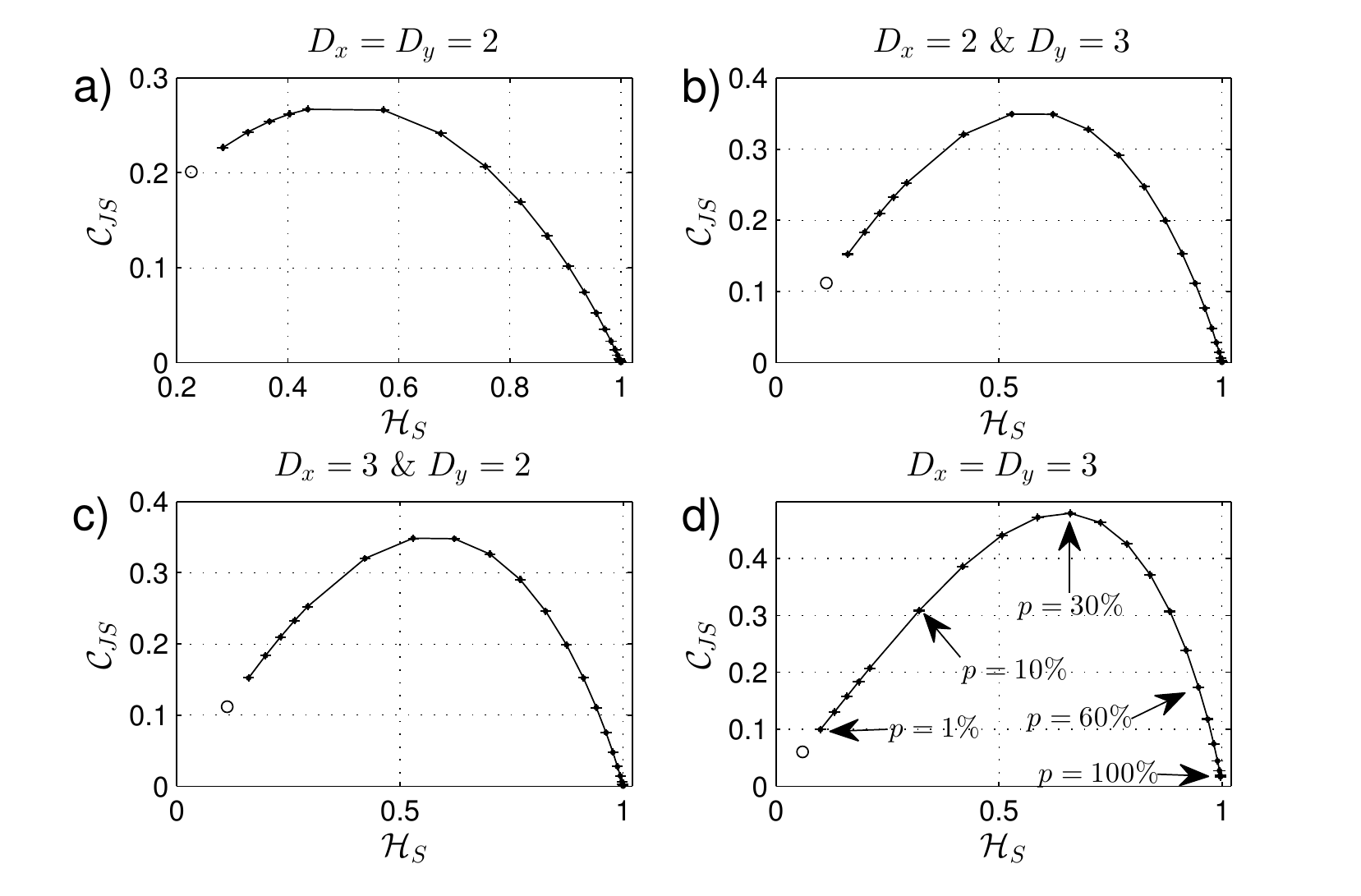}}
\caption{Curves described by the noisy periodic ornaments in the 2D CECP for different embedding dimensions: a) $D_x=D_y=2$, b) $D_x=2$ $\&$ $D_y=3$, c) $D_x=3$ $\&$ $D_y=2$, and d) $D_x=D_y=3$. The probability of change $p \in \{1,2,3,4,5,10,15,\dots,95,100\}$ is increasing from left to right. Mean and standard deviation of the ordinal quantifiers estimations for ten independent realizations for each $p-$value are plotted. Empty circles indicate the location of the original periodic ornament.}
\label{figure2}
\end{center}
\end{figure}

We have applied the proposed approach to these noisy ornaments in order to follow how the symbolic quantifiers change as a function of the probability of change $p$ ($p \in \{1,2,3,4,5,10,15,\dots,95,100\}$). Curves described in the 2D CECP for different embedding dimensions $D_x$ and $D_y$ are depicted in Fig.~\ref{figure2}. In this case, we have analized the images with their original spatial resolutions, \textit{i.e.} with $\tau_x=\tau_y=1$. For each value of the probability $p$, ten independent realizations were generated. Mean and standard deviation of the ordinal quantifiers for this set of realizations are plotted in Fig.~\ref{figure2}. Comparing Figs.~\ref{figure2}~b)~and~c), we noticed that the curve described in the CECP is invariant under the rotation $D_x \to D_y$ and $D_y \to D_x$, a feature that reflects the image symmetry. The 2D CECP locates the noisy ornaments in accord with intuitive notions, \textit{i.e.} regular two-dimensional patterns have lower values for both entropy and complexity, whereas random structureless ornaments have entropies near one and complexity close to zero. In the intermediate generated ornaments, hidden order behind a randomness environment can be visually discriminated. These more complex structures are suitably characterized by the 2D CECP. More precisely, a maximum of the normalized Jensen-Shannon complexity is reached at the frontier between order and disorder as it is expected. Particularly, the maximum complexity value estimated for embedding dimensions $D_x=D_y=3$ corresponds to the noisy ornament with parameter $p=30\%$ depicted in Fig.~\ref{figure1}~c).

\subsection{Multiscale analysis: discriminating roughness at different scales}
\label{Multiscale_analysis}
It is widely recognized that natural images are complex structures with different textures at relatively small and large spatial scales. As stressed by Alvarez-Ramirez~\textit{et al.}~\cite{alvarez-ramirez2006}: \textit{rarely natural textures can be characterized with a single roughness parameter.} Indeed, the multiscale Hurst exponent has been proposed as an alternative for quantifying roughness at various scales~\cite{kaplan1995,kaplan1999}. The Hurst exponent is a parameter that control the roughness of the fractional Brownian motion (fBm) model. Mandelbrot introduced this self-similar stochastic process for simulating natural textures~\cite{mandelbrot1982}.

\begin{figure}
\begin{center}
\resizebox{1.00\columnwidth}{!}{\includegraphics{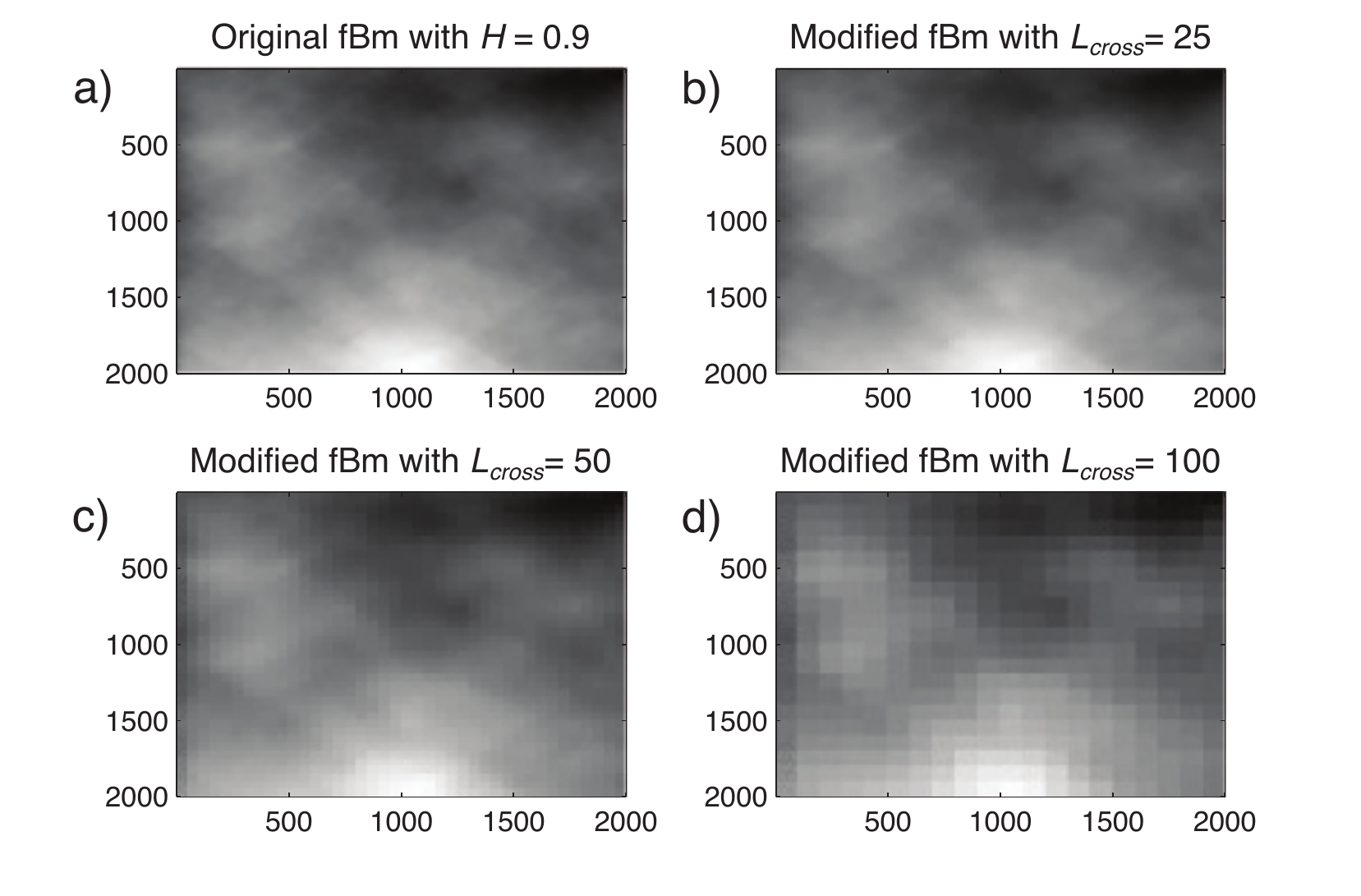}}
\caption{a) Numerically simulated fBm surface with Hurst exponent $H=0.9$ of size $2,000 \times 2,000$ pixels. b)-d) Modified fractal surfaces with $L_{cross}=25$, $L_{cross}=50$, and $L_{cross}=100$, respectively. Spatial correlations for scales lower than the crossover length $L_{cross}$ are destroyed by shuffling (randomly reordering).}
\label{figure3}
\end{center}
\end{figure}

\begin{figure}
\begin{center}
\resizebox{1.00\columnwidth}{!}{\includegraphics{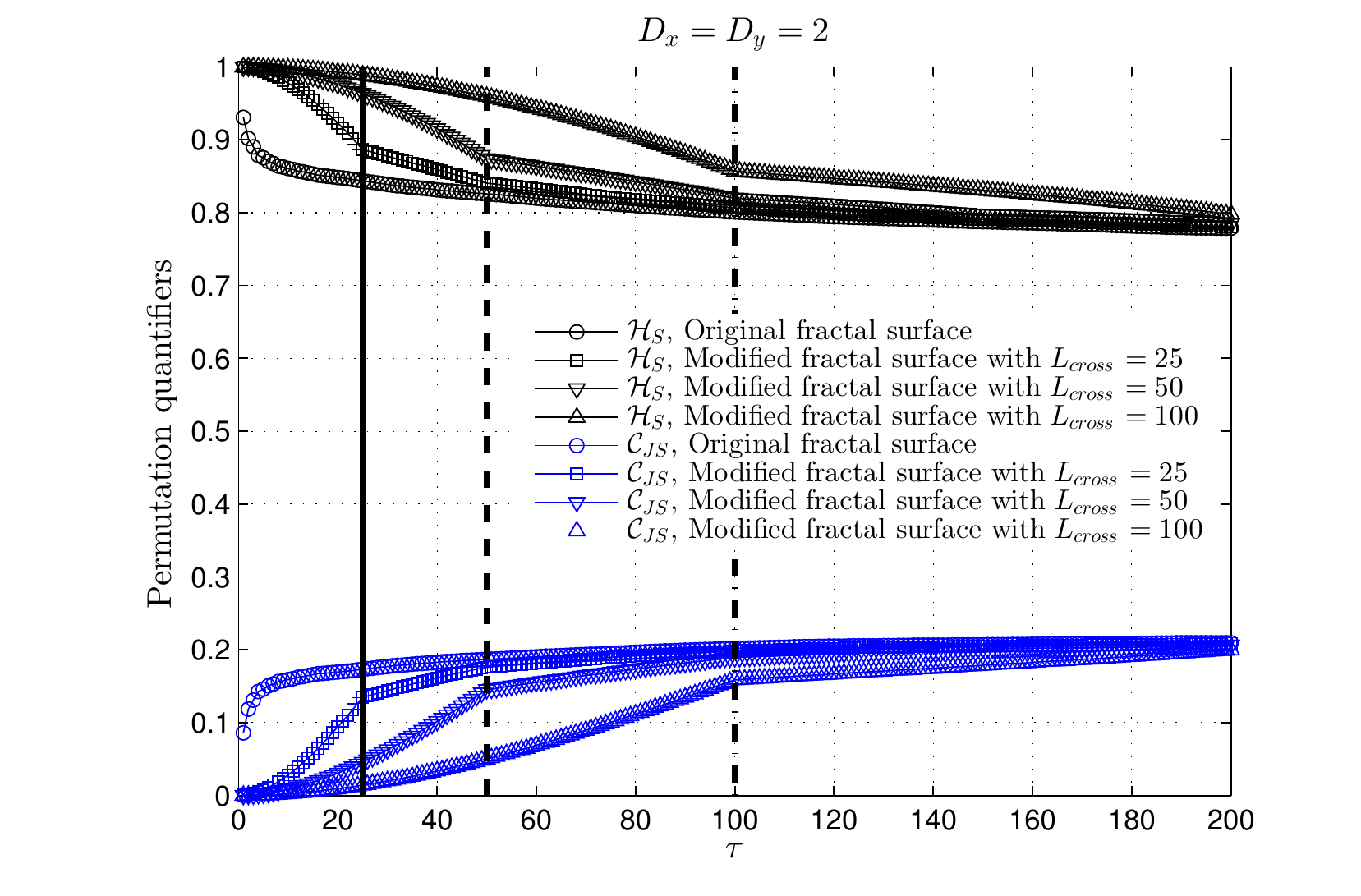}}
\caption{Estimated $\mathcal{H}_S$ and $\mathcal{C}_{JS}$ as a function of the embedding delay $\tau$ ($\tau_x=\tau_y=\tau$) with $D_x=D_y=2$ for the four images displayed in Fig.~\ref{figure3}. Vertical solid, dashed and dash-dotted black lines indicate the crossover lengths 25, 50 and 100, respectively.}
\label{figure4}
\end{center}
\end{figure}

In order to test the ability of the ordinal symbolic quantifiers for distinguishing between different textures at small and large resolution scales, a second numerically generated application is studied. A fBm surface with Hurst exponent $H=0.9$ is simulated through the random midpoint displacement algorithm. Further details about this method for generating fractal landscapes can be found in Ref.~\cite{ribeiro2012}. The simulated fractal surface, of size $2,000 \times 2,000$ pixels, is shown in Fig.~\ref{figure3}~a). The original image is then segmented into non-overlapping arrays of size $L_{cross} \times L_{cross}$ pixels, and the elements inside these arrays are subsequently shuffled spatially. In such a way, all the underlying spatial correlations for scales lower than the crossover length $L_{cross}$ are destroyed whereas spatial correlations at larger scales are conserved. Illustrative examples of the transformed images obtained by applying this procedure are depicted in Figs.~\ref{figure3}~b)-d) for crossover lengths equal to 25, 50, and 100, respectively. Both ordinal quantifiers ($\mathcal{H}_S$ \& $\mathcal{C}_{JS}$) are estimated for these four fractal surfaces fixing the embedding dimensions ($D_x=D_y=2$) and varying the embedding delays $\tau_x=\tau_y=\tau$ ($1 \leq \tau \leq 200$). Results are plotted in Fig.~\ref{figure4}. For low $\tau-$values, $\mathcal{H}_S$ and $\mathcal{C}_{JS}$ are close to 1 and 0, respectively, confirming the presence of uncorrelated random behavior for small spatial scales. On the other hand, the estimated values for both ordinal quantifiers tend to those associated with the original ``pure'' fBm surface with $H=0.9$ for large spatial scales. Thus, the proposed multiscale quantifiers are shown to be able to discriminate between different roughness features according to the considered spatial resolution. Last but not least, this analysis also confirms that the crossover length can be suitably identified with this approach.

\subsection{Normalized Brodatz texture database}
\label{Brodatz}
\begin{figure}
\begin{center}
\resizebox{1.00\columnwidth}{!}{\includegraphics{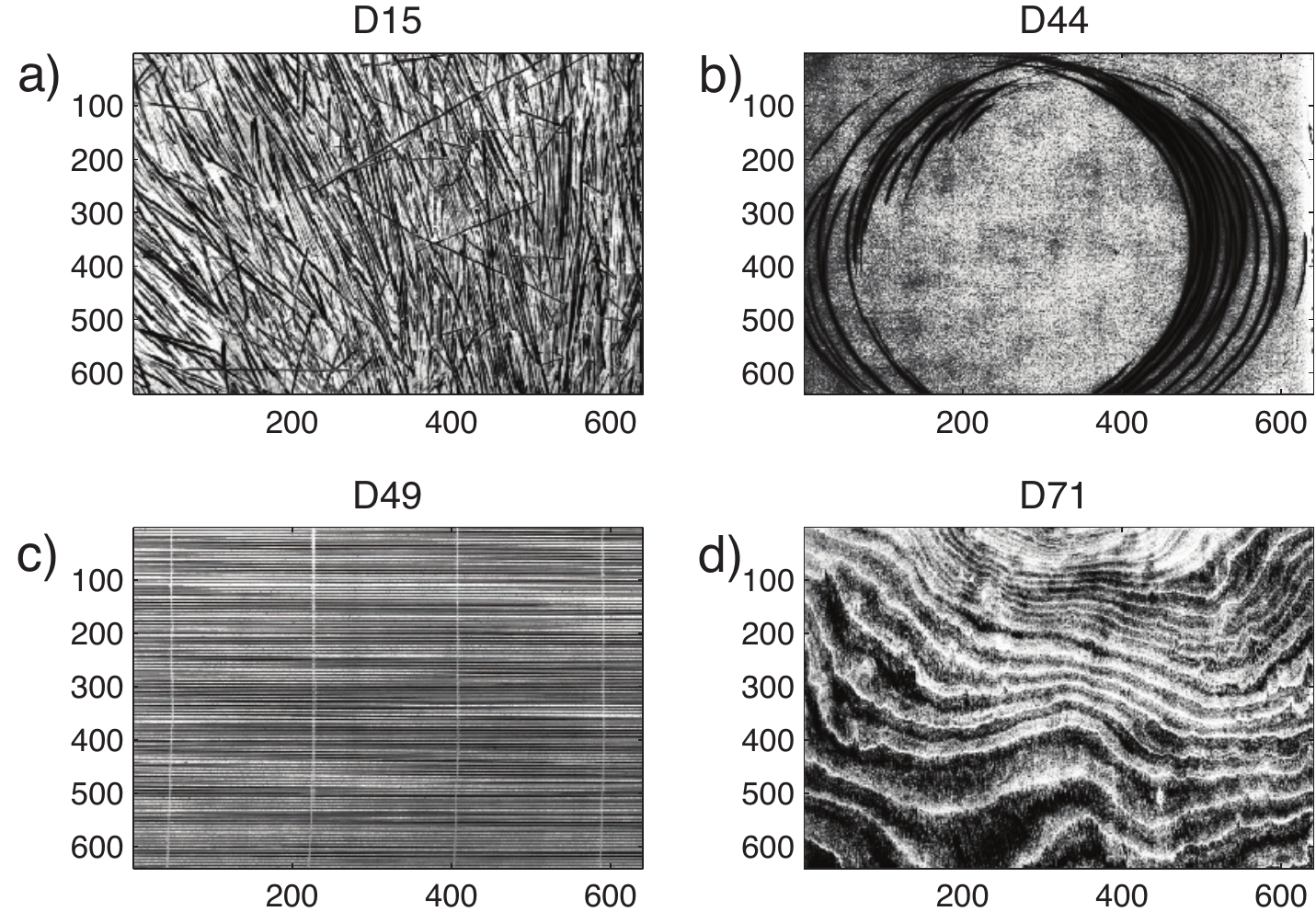}}
\caption{Four samples of the NBT album with their corresponding labels are illustrated: a) D15, b) D44, c) D49, and d) D71.}
\label{figure5}
\end{center}
\end{figure}

\begin{figure}
\begin{center}
\resizebox{1.00\columnwidth}{!}{\includegraphics{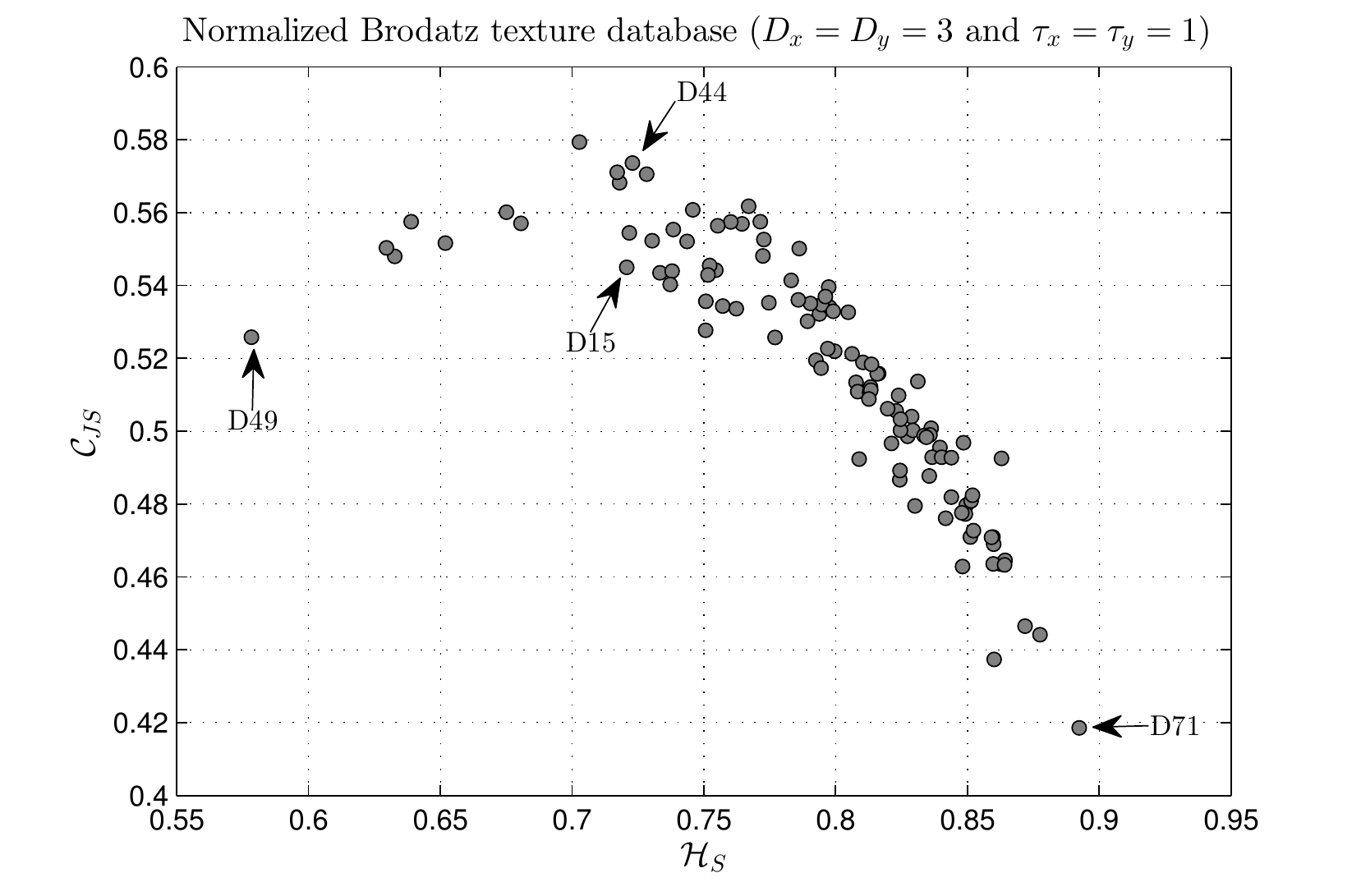}}
\caption{Location of the 112 texture images from the NBT album in the CECP. Quantifiers were estimated by implementing a symbolic reconstruction with $D_x=D_y=3$ and $\tau_x=\tau_y=1$.}
\label{figure6}
\end{center}
\end{figure}

The 112 texture images given in the Normalized Brodatz Texture (NBT) album have been analyzed through the 2D CECP. This normalized database is an improvement regarding the original Brodatz texture database since grayscale background effects have been removed~\cite{safia2013}. Consequently, it is impossible to discriminate between textures from this normalized database using only first order statistics. Four different samples of the NBT database are illustrated in Fig.~\ref{figure5}. It is worth remarking that the standard Brodatz grayscale texture album~\cite{brodatz1966}, composed of 112 grayscale images representing a large variety of natural textures, has been widely used as a validation dataset of texture methods (\textit{e.g.}, Refs.~\cite{sarkar1995,lee2010,florindo2012,goncalves2014,oliveira2015,davarzani2015}, among many others). The NBT image database is available at the following link: http://multibandtexture.recherche.usherbrooke.ca/normalized\_brodatz.html. The images of the NBT album have dimensions of $640 \times 640$ pixels and 8 bits/pixel, which provides 256 grayscale levels. Locations of these normalized 112 texture images in the CECP for $D_x=D_y=3$ and $\tau_x=\tau_y=1$ are shown in Fig.~\ref{figure6}. Notice that they spread over the proposed representation space. Locations of the four particular normalized Brodatz images displayed in Fig.~\ref{figure5} are indicated. It is worth noting that both quantifiers should be estimated to discriminate between the different texture images. For instance, D15 and D44 have approximately the same estimated $\mathcal{H}_S$ value. However, the calculated $\mathcal{C}_{JS}$ value for D44 is higher than that estimated for D15.

\begin{figure}
\begin{center}
\resizebox{0.90\columnwidth}{!}{\includegraphics{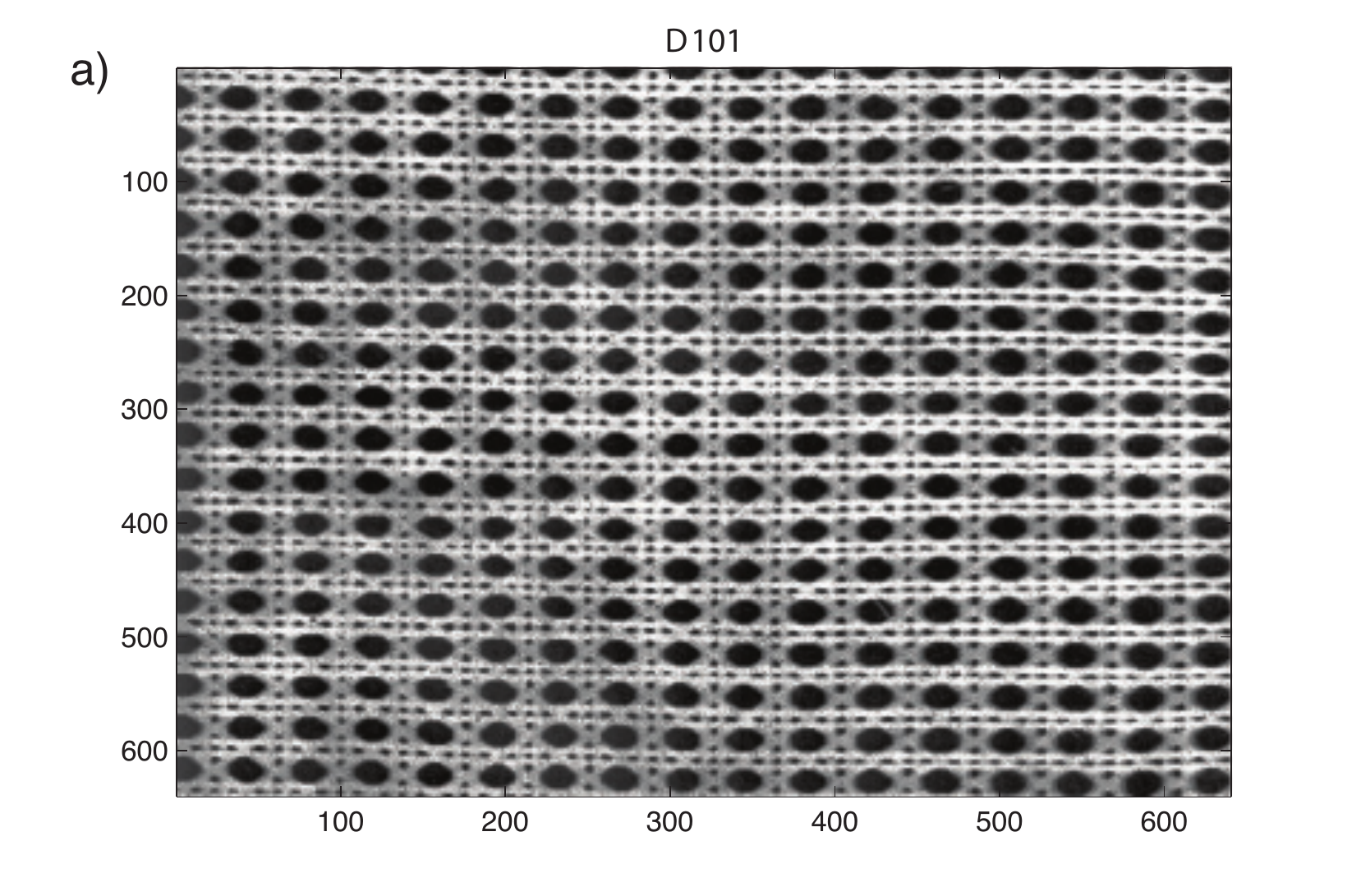}}
\resizebox{0.90\columnwidth}{!}{\includegraphics{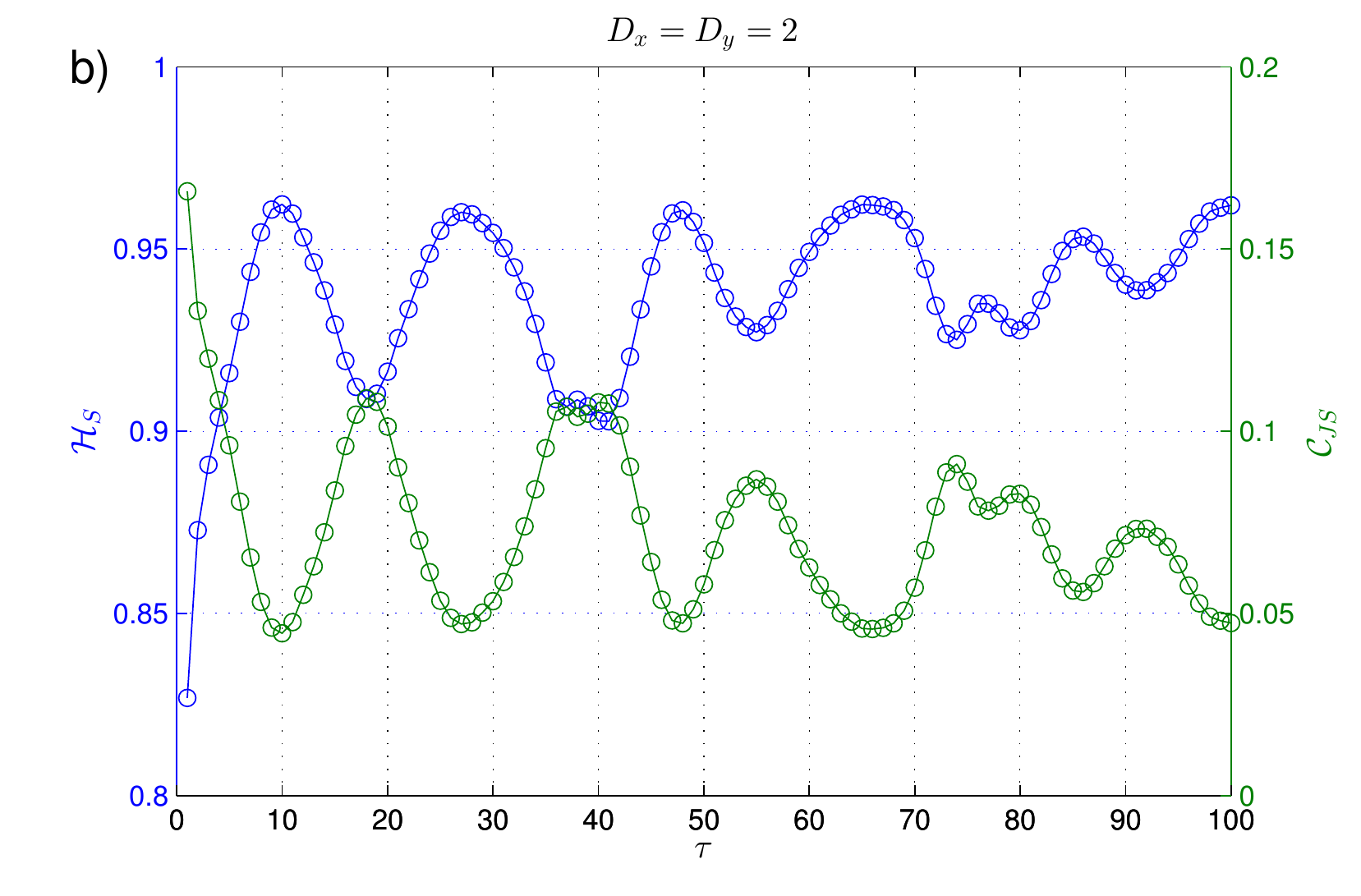}}
\caption{a) Normalized Brodatz texture labeled as D101. b) Multiscale analysis ($\tau_x=\tau_y=\tau, 1 \leq \tau \leq 100$) of this normalized Brodatz texture for embedding dimensions $D_x=D_y=2$.}
\label{figure7}
\end{center}
\end{figure}

Evidently, due to their intrinsic heterogeneity, texture images from the NBT database call for a multiscale analysis in order to achieve a more comprehensive characterization of their features at different scales. This contrasts with the homogeneity that characterizes self-similar structures (please compare with Fig.~\ref{figure3}~a)). As an illustrative example of this fact, we have performed the multiscale analysis of the normalized Brodatz texture labeled as D101 depicted in Fig.~\ref{figure7}~a). Permutation quantifiers as a function of the embedding delay $\tau$ ($\tau_x=\tau_y=\tau, 1 \leq \tau \leq 100$) for embedding dimensions $D_x=D_y=2$ are plotted in Fig.~\ref{figure7}~b). We noticed that the values estimated for both ordinal measures are strongly dependent on the spatial scale under which the image is analyzed. Furthermore, the underlying periodicity of the texture image is clearly reflected on the quantifiers' behaviors.

\subsection{Experimental applications}
\label{Experimental}
\begin{figure}
\begin{center}
\resizebox{1.00\columnwidth}{!}{\includegraphics{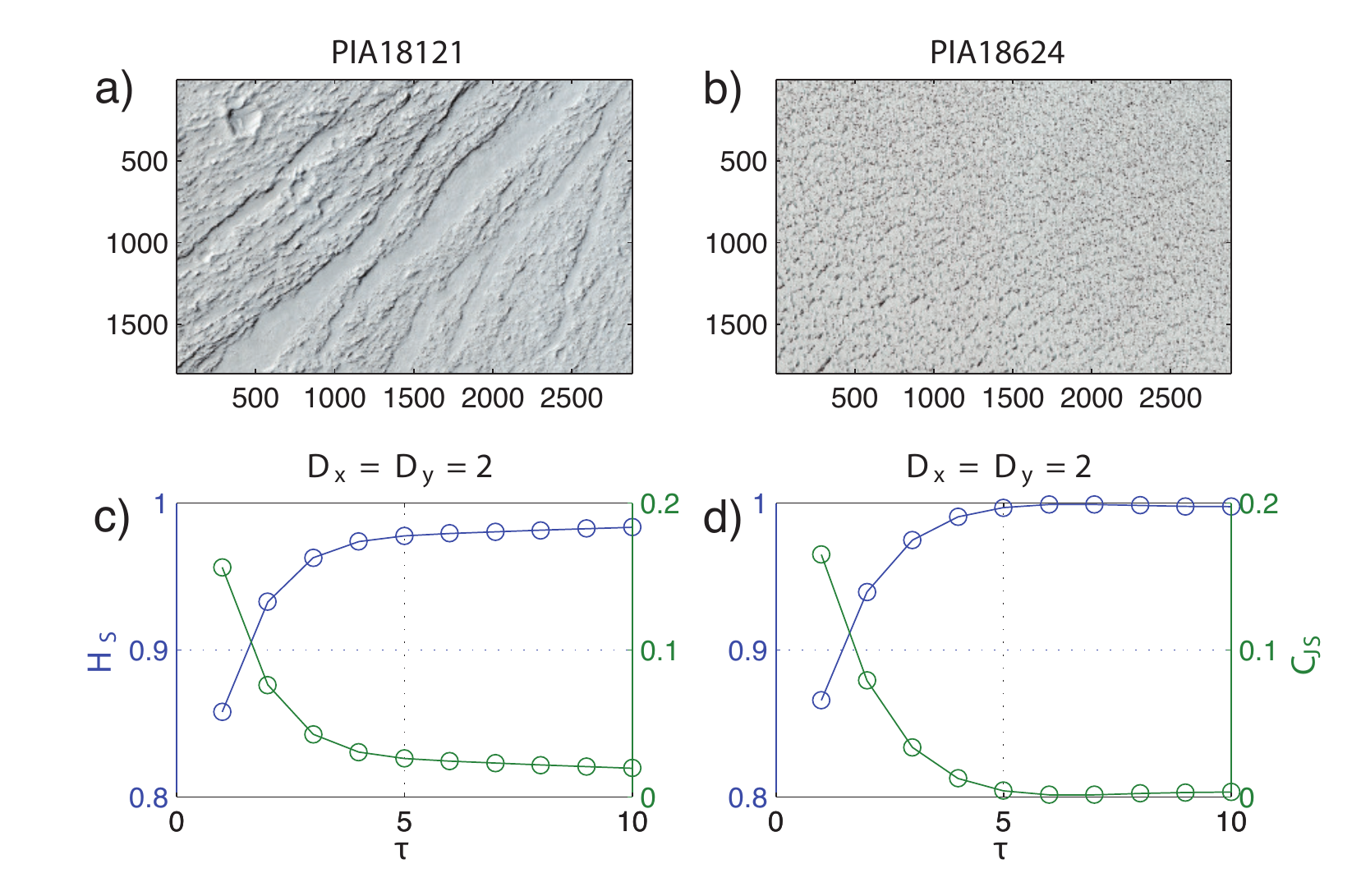}}
\caption{Real image analysis: a)-b) two high-resolution images from Mars (PIA18121 and PIA18624 from the Planetary Image Archive, Photojournal interface); c)-d) Multiscale analysis of the green channel performed for $D_x=D_y=2$, $\tau_x=\tau_y=\tau$ and $1 \leq \tau \leq 10$. Similar curves are obtained for the red and blue channels.}
\label{figure8}
\end{center}
\end{figure}

To test the performance of the multiscale approach in a practical context, we include here the analysis of real images. Two high-resolution images from Mars, publicly available in the Photojournal interface to the Planetary Image Archive (PIA) (Courtesy NASA/JPL-Caltech), have been carefully studied. More precisely, we have selected images labeled as PIA18121 (Chevrons on a Flow Surface in Marte Vallis, 2880 x 1800 pixels) and PIA18624 (The Icy Surface of the North Polar Cap, 2880 x 1800 pixels). They are displayed in Figs.~\ref{figure8}~a)~and~b), respectively. Further details about them can be obtained at the following link: http://photojournal.jpl.nasa.gov/. Permutation quantifiers were estimated for different embedding delays ($\tau_x=\tau_y=\tau$ and $1 \leq \tau \leq 10$) with $D_x=D_y=2$. Results obtained for the green channel of both images are shown in Figs.~\ref{figure8}~c)~and~d). Similar behaviors are found for the red and blue channels. For the original spatial sampling scale ($\tau=1$) these images have a similar texture according to the values estimated for $\mathcal{H}_S$ and $\mathcal{C}_{JS}$. However, their roughness properties can be distinguished for larger spatial scales since PIA18624 appears to be more random ($\mathcal{H}_S \to 1$ \& $\mathcal{C}_{JS} \to 0$) than PIA18121 for $\tau \geq 5$.

\begin{figure}
\begin{center}
\resizebox{0.70\columnwidth}{!}{\includegraphics{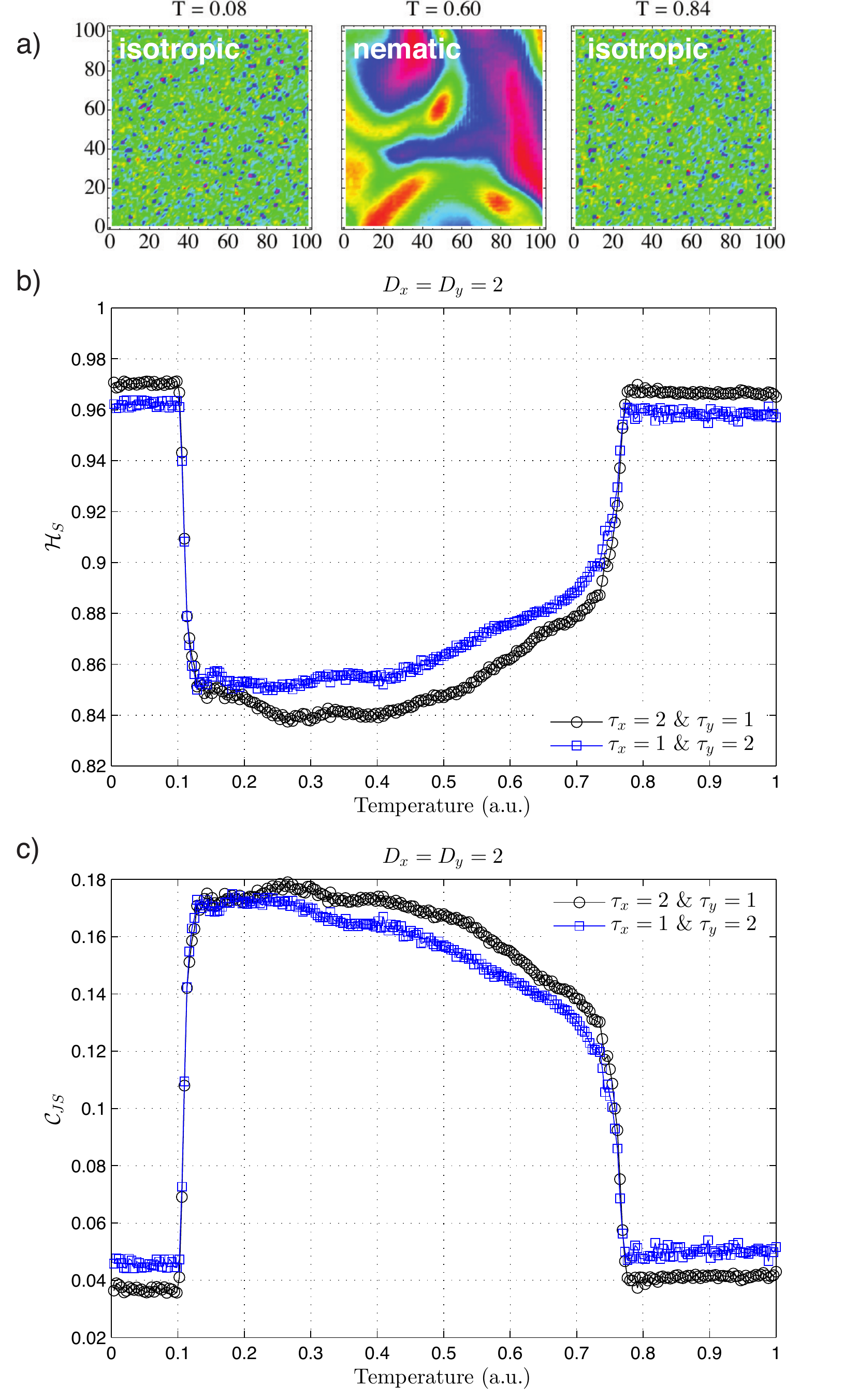}}
\caption{a) Examples of textures of a lyotropic liquid crystal in the (reentrant) isotropic, nematic and isotropic phases. Estimated values for b) $\mathcal{H}_S$ and c) $\mathcal{C}_{JS}$ as a function of the temperature with embedding dimensions $D_x=D_y=2$ and embedding delays $\tau_x=2$ and $\tau_y=1$ as well as $\tau_x=1$ and $\tau_y=2$.}
\label{figure9}
\end{center}
\end{figure}

Another experimental application we have studied is related to liquid crystal. In particular, we investigated the isotropic--nematic--isotropic phase transition of a thin sample of a lyotropic liquid crystal. These transition can be visualized by looking at the texture patterns of the sample in a polarized light microscope as function of the temperature. Figure~\ref{figure9}~a) shows three examples of these textures for three different temperatures. It is very clear that the isotropic phases are characterized by a noisy texture; on the other hand, the nematic phase displays a more complex pattern. The textures analyzed here are the same we have previously investigated in Ref.~\cite{ribeiro2012}. The values of $\mathcal{H}_S$ and $\mathcal{C}_{JS}$ were calculated considering $D_x=D_y=2$ and the embedding delays $\tau_x=2$ and $\tau_y=1$ as well as the $\tau_x=1$ and $\tau_y=2$ (that is, the rotated version); also, we have considered the average value of the pixels of the three layers (RGB) of the original images. As shown in Figs.~\ref{figure9}~b)~and~c), the permutation quantifiers are able to identify the transitions from isotropic to nematic and from nematic to isotropic, a feature that is actually not surprising due to the very distinct patterns of the textures in these mesophases. Intriguingly, we observe that the values of $\mathcal{H}_S$ in the isotropic phases for $\tau_x=1$ and $\tau_y=2$ are systematically smaller than the values obtained for $\tau_x=2$ and $\tau_y=1$. Reciprocally, values of $\mathcal{C}_{JS}$ are larger for $\tau_x=1$ and $\tau_y=2$ than for $\tau_x=2$ and $\tau_y=1$. Thus, these results suggest that there are more ordered structures along the $y$ direction, which matches the axis of the elongated capillary tube where the sample was placed. Because of this coincidence, it is very hard to not infer that this ``additional'' order is caused by a surface effect acting on liquid crystal molecules. It is well known that, despite of the isotropic phase been macroscopically homogeneous, locally (on a short distance scale) nematic-like order exists and can be caused by a surface effect~\cite{khoo2007}. Here, despite the lack of more experimental evidence, we believe that our multiscale generalization may be employed as a very simple tool for identifying local order in isotropic phases, a task that is usually more tricky~\cite{khoo2007}.

\section{Conclusions}
\label{Conclusions}
In this study, we address the problem of the identification and characterization of spatial correlations in two-dimensional patterns. The proposed multiscale 2D CECP provides a feasible alternative for the quantitative characterization and discrimination of two-dimensional structures at different spatial scales from a novel, information-theory, perspective. It could complement the information extracted from more traditional tools. Taking into account the key role played by texture analysis in the image processing from a wide number of significant applications, including remote sensing, assisted medical diagnosis, and automatic target recognition, we consider that this multiscale symbolic approach can be of potential interest and utility to researchers working in these fields. Moreover, thanks to its robustness to noise, the multiscale 2D CECP seems to be especially suited for the analysis of experimental images. In particular, we conjecture that this methodology could be successfully applied in the digital image encryption field to objectively quantify the efficiency of different image encryption schemes. We are planning to address this hypothesis in a future study. The possibility to replace the original quantifiers, namely permutation entropy and permutation statistical complexity, by other alternatives, such as the cluster entropy~\cite{carbone2004,carbone2013}, looking for potential improvements could be another avenue of research for the future.

\section*{Acknowledgements}
LZ acknowledges Consejo Nacional de Investigaciones Cient\'ificas y T\'ecnicas (CONICET), Argentina and Universidad Nacional de La Plata, Argentina (Incentive Project 11/I170) for their financial support. HVR thanks the financial support of the CNPq under Grant No. 440650/2014-3.

\section*{References}
\bibliographystyle{elsarticle-num}
\bibliography{CECP_2D}

\begin{thebibliography}{10}
\expandafter\ifx\csname url\endcsname\relax
  \def\url#1{\texttt{#1}}\fi
\expandafter\ifx\csname urlprefix\endcsname\relax\def\urlprefix{URL }\fi
\expandafter\ifx\csname href\endcsname\relax
  \def\href#1#2{#2} \def\path#1{#1}\fi

\bibitem{feldman2003}
D.~P. Feldman, J.~P. Crutchfield, Structural information in two-dimensional
  patterns: {Entropy} convergence and excess entropy, Phys. Rev. E 67~(5)
  (2003) 051104.

\bibitem{andrienko2000}
Y.~A. Andrienko, N.~V. Brilliantov, J.~Kurths, Complexity of two-dimensional
  patterns, Eur. Phys. J. B 15~(3) (2000) 539--546.

\bibitem{cai2006}
Z.~Cai, E.~Shen, F.~Gu, Z.~Xu, J.~Ruan, Y.~Cao, A new two-dimensional
  complexity measure, Int. J. Bifurcation Chaos 16~(11) (2006) 3235--3247.

\bibitem{kaplan1995}
L.~M. Kaplan, C.-C.~J. Kuo, Texture roughness analysis and synthesis via
  extended self-similar ({ESS}) model, IEEE Trans. Pattern Anal. Machine
  Intell. 17~(11) (1995) 1043--1056.

\bibitem{kaplan1999}
L.~M. Kaplan, Extended fractal analysis for texture classification and
  segmentation, IEEE Trans. Image Processing 8~(11) (1999) 1572--1585.

\bibitem{alvarez-ramirez2006}
J.~Alvarez-Ramirez, E.~Rodriguez, I.~Cervantes, J.~C. Echeverria, Scaling
  properties of image textures: {A} detrending fluctuation analysis approach,
  Physica A 361~(2) (2006) 677--698.

\bibitem{gu2006}
G.-F. Gu, W.-X. Zhou, Detrended fluctuation analysis for fractals and
  multifractals in higher dimensions, Phys. Rev. E 74~(6) (2006) 061104.

\bibitem{carbone2007}
A.~Carbone, Algorithm to estimate the {Hurst} exponent of high-dimensional
  fractals, Phys. Rev. E 76~(5) (2007) 056703.

\bibitem{mandelbrot1982}
B.~B. Mandelbrot, The fractal geometry of nature, W. H. Freeman and Company,
  1982.

\bibitem{wang2013}
F.~Wang, J.-W. Li, W.~Shi, G.-P. Liao, Leaf image segmentation method based on
  multifractal detrended fluctuation analysis, J. Appl. Phys. 114~(21) (2013)
  214905.

\bibitem{valdiviezo2014}
J.~C. {Valdiviezo-N.}, R.~Castro, G.~Crist\'obal, A.~Carbone, Hurst exponent
  for fractal characterization of {LANDSAT} images, SPIE Proceedings 9221
  (2014) 922103.

\bibitem{vargas2015}
C.~Vargas-Olmos, J.~S. Murgu\'ia, M.~T. Ram\'irez-Torres, M.~{Mej\'ia Carlos},
  H.~C. Rosu, H.~Gonz\'alez-Aguilar, Two-dimensional {DFA} scaling analysis
  applied to encrypted images, Int. J. Mod. Phys. C 26~(8) (2015) 1550093.

\bibitem{ribeiro2012}
H.~V. Ribeiro, L.~Zunino, E.~K. Lenzi, P.~A. Santoro, R.~S. Mendes,
  Complexity-entropy causality plane as a complexity measure for
  two-dimensional patterns, PLoS ONE 7~(8) (2012) e40689.

\bibitem{bandt2002}
C.~Bandt, B.~Pompe, Permutation entropy: {A} natural complexity measure for
  time series, Phys. Rev. Lett. 88~(17) (2002) 174102.

\bibitem{rosso2007}
O.~A. Rosso, H.~A. Larrondo, M.~T. Martin, A.~Plastino, M.~A. Fuentes,
  Distinguishing noise from chaos, Phys. Rev. Lett. 99~(15) (2007) 154102.

\bibitem{feldman1998}
D.~P. Feldman, J.~P. Crutchfield, Measures of statistical complexity: {Why?},
  Phys. Lett. A 238~(4-5) (1998) 244--252.

\bibitem{lange2013}
H.~Lange, O.~A. Rosso, M.~Hauhs, Ordinal pattern and statistical complexity
  analysis of daily stream flow time series, Eur. Phys. J. Special Topics
  222~(2) (2013) 535--552.

\bibitem{wackerbauer1994}
R.~Wackerbauer, A.~Witt, H.~Atmanspacher, J.~Kurths, H.~Scheingraber, A
  comparative classification of complexity measures, Chaos, Solitons \&
  Fractals 4~(1) (1994) 133--173.

\bibitem{lamberti2004}
P.~W. Lamberti, M.~T. Martin, A.~Plastino, O.~A. Rosso, Intensive entropic
  non-triviality measure, Physica A 334~(1-2) (2004) 119--131.

\bibitem{lopezruiz1995}
R.~L\'opez-Ruiz, H.~L. Mancini, X.~Calbet, A statistical measure of complexity,
  Phys. Lett. A 209~(5-6) (1995) 321--326.

\bibitem{Lin1991}
J.~Lin, Divergence measures based on the {Shannon} entropy, IEEE Trans. Inf.
  Theory 37~(1) (1991) 145--151.

\bibitem{grosse2002}
I.~Grosse, P.~Bernaola-Galv\'an, P.~Carpena, R.~Rom\'an-Rold\'an, J.~Oliver,
  H.~E. Stanley, Analysis of symbolic sequences using the {Jensen-Shannon}
  divergence, Phys. Rev. E 65~(4) (2002) 041905.

\bibitem{martin2006}
M.~T. Martin, A.~Plastino, O.~A. Rosso, Generalized statistical complexity
  measures: {Geometrical} and analytical properties, Physica A 369~(2) (2006)
  439--462.

\bibitem{zunino2012}
L.~Zunino, M.~C. Soriano, O.~A. Rosso, Distinguishing chaotic and stochastic
  dynamics from time series by using a multiscale symbolic approach, Phys. Rev.
  E 86~(4) (2012) 046210.

\bibitem{rosso2009}
O.~A. Rosso, H.~Craig, P.~Moscato, Shakespeare and other {English Renaissance}
  authors as characterized by {Information Theory} complexity quantifiers,
  Physica A 388~(6) (2009) 916--926.

\bibitem{zunino2010}
L.~Zunino, M.~Zanin, B.~M. Tabak, D.~G. P\'erez, O.~A. Rosso,
  Complexity-entropy causality plane: {A} useful approach to quantify the stock
  market inefficiency, Physica A 389~(9) (2010) 1891--1901.

\bibitem{zunino2011}
L.~Zunino, B.~M. Tabak, F.~Serinaldi, M.~Zanin, D.~G. P\'erez, O.~A. Rosso,
  Commodity predictability analysis with a permutation information theory
  approach, Physica A 390~(5) (2011) 876--890.

\bibitem{zunino2012b}
L.~Zunino, A.~F. Bariviera, M.~B. Guercio, L.~B. Martinez, O.~A. Rosso, On the
  efficiency of sovereign bond markets, Physica 391~(18) (2012) 4342--4349.

\bibitem{fernandez_bariviera2013}
A.~F. Bariviera, L.~Zunino, M.~B. Guercio, L.~B. Martinez, O.~A. Rosso,
  Efficiency and credit ratings: a permutation-information-theory analysis, J.
  Stat. Mech. (2013) P08007.

\bibitem{ribeiro2012b}
H.~V. Ribeiro, L.~Zunino, R.~S. Mendes, E.~K. Lenzi, Complexity-entropy
  causality plane: {A} useful approach for distinguishing songs, Physica A
  391~(7) (2012) 2421--2428.

\bibitem{lovallo2011}
M.~Lovallo, L.~Telesca, Complexity measures and information planes of x-ray
  astrophysical sources, J. Stat. Mech. (2011) P03029.

\bibitem{serinaldi2014}
F.~Serinaldi, L.~Zunino, O.~A. Rosso, Complexity-entropy analysis of daily
  stream flow time series in the continental {United States}, Stochastic
  Environmental Research and Risk Assessment 28~(7) (2014) 1685--1708.

\bibitem{montani2014}
F.~Montani, O.~A. Rosso, Entropy-complexity characterization of brain
  development in chickens, Entropy 16~(8) (2014) 4677--4692.

\bibitem{li2014}
Q.~Li, F.~Zuntao, Permutation entropy and statistical complexity quantifier of
  nonstationarity effect in the vertical velocity records, Phys. Rev. E 89~(1)
  (2014) 012905.

\bibitem{weck2015}
P.~J. Weck, D.~A. Schaffner, M.~R. Brown, R.~T. Wicks, Permutation entropy and
  statistical complexity analysis of turbulence in laboratory plasmas and the
  solar wind, Phys. Rev. E 91~(2) (2015) 023101.

\bibitem{safia2013}
A.~Safia, D.-C. He, New {Brodatz}-based image databases for grayscale color and
  multiband texture analysis, ISRN Machine Vision 2013 (2013) 876386.

\bibitem{brodatz1966}
P.~Brodatz, Textures: A Photographic Album for Artist \& Designers, Dover, New
  York, USA, 1966.

\bibitem{sarkar1995}
N.~Sarkar, B.~B. Chaudhuri, Multifractal and generalized dimensions of
  gray-tone digital images, Signal Processing 42~(2) (1995) 181--190.

\bibitem{lee2010}
W.-L. Lee, K.-S. Hsieh, A robust algorithm for the fractal dimension of images
  and its applications to the classification of natural images and ultrasonic
  liver images, Signal Processing 90~(6) (2010) 1894--1904.

\bibitem{florindo2012}
J.~B. Florindo, O.~M. Bruno, Fractal descriptors based on {Fourier} spectrum
  applied to texture analysis, Physica A 391~(20) (2012) 4909--4922.

\bibitem{goncalves2014}
W.~N. Gon{\c{c}}alves, B.~B. Machado, O.~M. Bruno, Texture descriptor combining
  fractal dimension and artificial crawlers, Physica 395 (2014) 358--370.

\bibitem{oliveira2015}
M.~W. da~Silva~Oliveira, N.~R. da~Silva, A.~Manzanera, O.~M. Bruno, Feature
  extraction on local jet space for texture classification, Physica A 439
  (2015) 160--170.

\bibitem{davarzani2015}
R.~Davarzani, S.~Mozaffari, K.~Yaghmaie, Scale- and rotation-invariant texture
  description with improved local binary pattern features, Signal Processing
  111 (2015) 274--293.

\bibitem{khoo2007}
I.-C. Khoo, Liquid Crystals, 2nd Edition, Wiley Series in Pure and Applied
  Optics, Wiley, 2007.

\bibitem{carbone2004}
A.~Carbone, H.~E. Stanley, Scaling properties and entropy of long-range
  correlated time series, Physica A 384~(1) (2004) 21--24.

\bibitem{carbone2013}
A.~Carbone, Information measure for long-range correlated sequences: the case
  of the 24 human chromosomes, Scientific Reports 3 (2013) 2721.

\end{thebibliography}
\end{document}